\documentclass[epj]{svjour}
\usepackage{graphics}
 \usepackage{psfig,epsfig}
\newcommand{\ga}{\alpha}
\newcommand{\gb}{\beta}
\newcommand{\gc}{\gamma}
\newcommand{\gd}{\delta}
\newcommand{\gk}{\kappa}

\newcommand{\gs}{\sigma}

\begin{document}
\title{Light clusters in nuclear matter of finite temperature}
\author{M. Beyer, S. Strauss\inst{1} \and P. Schuck\inst{2} \and
  S.A. Sofianos\inst{3}
}                     
\institute{Fachbereich Physik, Universit\"at Rostock, D-18051 Rostock, Germany
\and Institut de Physique Nucl\'eaire, F-91406, Orsay Cedex, France
\and Physics Department, University of South Africa, Pretoria 0003,
  South Africa}
\date{Received: date / Revised version: date}
%
\abstract{
We investigate properties and the distribution of light nuclei ($A\le4$) in
  symmetric nuclear matter of finite temperature within a microscopic
  framework.  For this purpose we have solved few-body Alt-Grassberger-Sandhas
  type equations for quasi-nucleons that include self-energy corrections and
  Pauli blocking in a systematic way. In a statistical model we find a
  significant influence in the composition of nuclear matter if medium effects
  are included in the microscopic calculation of nuclei.  If multiplicities
  are frozen out at a certain time (or volume), we expect significant
  consequences for the formation of light fragments in a heavy ion collision.
  As a consequence of the systematic inclusion of medium effects the ordering
  of multiplicities becomes opposite to the law of mass action of ideal
  components.  This is necessary to explain the large abundance of
  $\alpha$-particles in a heavy ion collision that are otherwise largely
  suppressed in an ideal equilibrium scenario.
} 
\maketitle

\section{Introduction}

Heavy ion collisions provide a tool to investigate the phase structure of
nuclear matter. Depending on the energies, the region of temperature and
density explored might be rather large. The information about the composition
of nuclear matter is contained in the equation of state.  At collision
energies per nucleon well below one GeV the equation of state is described by
purely hadronic degrees of freedom.  It is a basic ingredient in microscopic
simulations of the heavy ion collision, such as the
Boltzmann-Uehling-Uhlenbeck (BUU)~\cite{dan91,dan92,sto86,fuc95,Beyer:1999xv}
or the quantum molecular dynamics (QMD)~\cite{aic91,pei92,neb99} simulations.
The challenge is to extract information about the different stages of the
evolution of the heavy ion collision.  This information could be provided by
fragments produced in the different stages of the collision as has been done
recently for the case $^{129}$Xe+$^\mathrm{nat}$Sn by the INDRA
collaboration~\cite{INDRA00,INDRA02}.

An early analysis of multi-fragmentation in a heavy ion collision of
$^{36}$Ar$+^{58}$Ni at several energies below $100A$ MeV has been given in
Ref.~\cite{Borderie:1995sk}. The authors study a class of evaporation events
at central collisions~\cite{Borderie:1995xc}.  During these events more than
90\% of the charged particles were detected and isotopically identified.
Within a thermal (and chemical) equilibrium scenario~\cite{mek78} of ideal
gas components (including states up to excited $^9$B), supplemented by finite
volume effects~\cite{tri94} and a model of side-feeding, they found a
remarkable agreement with the experimental data~\cite{Borderie:1995sk}.
Temperature has been varied between 10 and 25 MeV and the freeze-out volume
fixed to 1/3 of normal nuclear matter density.

A more elaborated statistical analysis has been done for the recent INDRA
experiment $^{129}$Xe+$^\mathrm{nat}$Sn. The measured multiplicities of the
central collisions by the INDRA collaboration show a large fraction of
$\alpha$-particles~\cite{INDRA02}. In contrast, a naive model of a gas of
ideal components, would give a much smaller number of $\alpha$-particles,
depending on the freeze-out density.  The INDRA collaboration provides a
detailed comparison of their data within a statistical multi-fragmentation
model (SMM)~\cite{bon95}. This model goes beyond a simple picture of an ideal
gas and describes multiplicities and some other aspects of the heavy ion
collision in question~\cite{INDRA02}.

On the other hand, from a microscopic analysis of cluster formation, it is
known that nuclei dissociate already at rather moderate densities and
temperatures, see, e.g.~\cite{Beyer:1999zx,Beyer:2000ds} and references
therein. Some details will also be given in this paper. The dissociation (Mott
effect) is taken into account, e.g., in modern BUU simulations of heavy ion
collisions and is necessary to reproduce the experimental data, see
Ref.~\cite{dan91,dan92}.  It is an effect related to the Pauli blocking
induced by the surrounding medium and goes beyond the picture of a simple
ideal gas of nuclei.

In a recent analysis of the central collision Xe+Sn at $50A$ MeV
that has been measured by the INDRA collaboration~\cite{INDRA00}, we found
that the BUU simulation gives a proton to deuteron ratio that is close to the
one expected by the equilibrium distribution; to be more precise, for times
$t>50$ fm/c during the evolution of the system. This however holds only, if the
equilibrium distribution includes the above mentioned dissociation of the
deuteron~\cite{Beyer:1999xv,ckuhrts}.

Therefore, we address the question to what extend the dissociation of nuclei
affects the equilibrium distribution of nuclear matter. To do so, we
investigate a system of light nuclei at finite density and temperature up to
the $\alpha$-particle.

Hence, we focus on a new aspect in the distribution of light nuclei.  Since
reasonable generalizations of the Feynman-Galitskii or Bethe-Goldstone
equations for more than two particles become available, the properties of
multi-particle correlations in a medium can now be addressed microscopically.
To demonstrate the effect that is related to in-medium properties of the light
clusters in question, in particular the $\alpha$-particle, we explore an {\em
  ab initio} equilibrium quantum statistical description of a many-particle
system based on the well established and successful (equilibrium) Green
function method~\cite{fet71}.  To include a proper description of clusters we
implement an equal time constraint on the Green functions.  This allows for a
cluster expansion of the Green functions as shown, for example
in~\cite{duk98}. In an uncorrelated medium of quasi-particles the equal time
constraint systematically leads to Dyson equations for clusters with a fixed
number of particles. They include the self energy corrections and the Pauli
blocking and are rearranged as resolved equations to use the
Alt-Grassberger-Sandhas (AGS) formalism to solve the respective few-body
equations which has been done for the three- and four-nucleon system
in~\cite{bey96,Beyer:1999zx,Beyer:2000ds}.  Similar equations to treat the
in-medium three-body system have been proposed previously in
Refs.~\cite{eic68,sch73}.  As an interaction we use a nucleon nucleon
potential that reasonably reproduces the nucleon nucleon phase shifts and the
binding energies of the light nuclei in question.

In Section~\ref{sec:stat} we introduce the consequences of the abovementioned
cluster expansion method in the equation of state.  This will be done along
the lines of~\cite{schulz}. The microscopic AGS-type equations to treat
multi-particle clusters in medium will be explained in Section~\ref{sec:AGS}.
We use a nucleon-nucleon potential that reproduces reasonably well the
nucleon-nucleon scattering data and the binding energies of light nuclei
considered. In Section~\ref{sec:results} we present our results. In
particular, we calculate the equilibrium composition of nuclear matter for
conditions comparable to the heavy ion collision investigated by the INDRA
collaboration. We summarize our conclusions in Section~\ref{sec:con}.

\section{Statistical model} \label{sec:stat}

To generalize the equation of state for a Fermi system that includes
correlations the nuclear matter density $n=n(\mu,T)$ as a function of the
chemical potential $\mu$ and temperature $T$ can be rearranged in an
uncorrelated part $n_{\rm free}$ and a correlated one $n_{\rm
  corr}$~\cite{schulz,Beyer:1999zx},
\begin{equation}
n=n_{\rm free}+n_{\rm corr}.
\end{equation}
To abbreviate notation let 1 denote the quantum numbers of particle 1.  The
Fermi function is given by
\begin{equation}
f(1)\equiv f(E_1)=\{\exp[\gb(E_1-\mu)]+1\}^{-1}
\end{equation}
where $E_1$ denotes the one-particle energy and $\gb$ the inverse temperature.
Presently we describe symmetric nuclear matter and hence
\begin{equation}
n_{\rm free}= 4\sum_1 f(1).
\end{equation}
For a system of nucleons of mass $m_N$ the energy is $E_1=k^2/2m_N$.
Hartree-Fock approximation introduces the notion of quasi-particles and
quasi-particle energies
\begin{equation}
E_1\rightarrow\varepsilon_1 = k^2/2m_N+\Sigma(k),
\label{eqn:quasi}
\end{equation}
where 
\begin{equation}
\Sigma(1) =\sum_{2}V_2(12,\widetilde{12})f(2).
\end{equation}
The tilde means proper anti-symmetrization (i.e. including the Fock term).

As explained in Ref.~\cite{Beyer:1999zx} in some detail, the correlated 
density can be composed into different cluster contributions.
\begin{equation}
n_{\mathrm{corr}}= 2n_2+3n_3+4n_4+\dots,\qquad
n_A=n_A^{\mathrm{b}}+n_A^{\mathrm{sc}}
\end{equation}
where $n_A$ denotes the $A$-particle correlated density presented as bound
$n_A^{\mathrm{b}}$ or scattering $n_A^{\mathrm{sc}}$ states in chemical
equilibrium. The full expression for the two-particle correlated densities
$n_2$ has been given in Ref.~\cite{schulz}.

To evaluate correlated densities we presently focus on the bound state
contributions. This is justified in view of the rather low densities of the
final stage of heavy ion collisions. Note, however, that including
contributions of scattering states requires major theoretical effort to
solve the respective scatting few-body equations derived in the next section.
The distribution functions for the $A$-body cluster of fermions is given by
\begin{equation}
f_A(p)=\left\{\exp[\gb(E_A-B_A-\mu_A)]+\epsilon\right\}^{-1}
\end{equation}
where $p$ is the c.m. momentum of the cluster, $E_A(p)$ is the continuum
energy, $B_A>0$ the binding energy of the cluster, $\epsilon=+1(-1)$ for
fermions (bosons), and $\mu_A$ the respective chemical potential. In
equilibrium considered here $\mu_A=A\mu$.  The density for the nucleus
of mass number $A$ is given by
\begin{equation}
n^\mathrm{b}_A(\mu,T) = (2S+1)(2I+1)\sum_p f_A(p)
\label{eqn:dens}
\end{equation}
where $S$ denotes the spin and $I$ the isospin of the nucleus.

\section{In-medium few-body
  equations} \label{sec:AGS}

The basis of the equations given in this section is the (equilibrium) Green
function approach to describe quantum statistical systems~\cite{fet71}. The
Green functions for a given number of particles are evaluated at equal
imaginary times assuming an environment of independent quasi-particles, see
e.g.~\cite{duk98}. As a consequence, for a given number of particles, a
Dyson-type equation can be derived that is only driven by the dynamics of the
smaller cluster (cluster mean field approximation) that breaks the Green
functions hierarchy. Utilizing resolvents the Dyson-type equation for a
particular cluster can be rewritten as AGS-type
equation~\cite{alt67,san74,alt72,san75} with an effective Hamiltonian.  This
has been shown previously for the nucleon deuteron reaction~\cite{bey96} the
three-nucleon bound state~\cite{Beyer:1999zx} and the
$\alpha$-particle~\cite{Beyer:2000ds}. In this section we briefly repeat the
relevant formulas to introduce our notation.

Utilizing the Dyson equation for clusters it is possible to introduce
resolvents to describe the dynamics of the system. Defining $H_0=\sum_{i=1}^n
\varepsilon_i$, with the quasi-particle self energy $\varepsilon_i$ the
$n$-quasi-particle cluster resolvent $G_0$ is
\begin{equation}
G_0(z) = (z-  H_0)^{-1}
\;{N} \equiv R_0(z)\;{ N}.
\end{equation}
Here $G_0$, $H_0$, and $N$ are matrices in $n$ particle space and $z$ denotes
the analytic continuation of the Matsubara frequency~\cite{fet71}. The
Pauli-blocking factors for $n$-particles are
\begin{equation}
N=\bar f(1)\bar f(2) \dots \bar f(n)
+\epsilon f(1)f(2)\dots f(n),
\end{equation}
with $\bar f\equiv 1-f$. Note: $NR_0=R_0N$.  Defining the effective potential
$V\equiv \sum_{\mathrm{pairs}\;\ga} { N_2^{\ga}}V_2^{\ga}$ the full, $G(z)$,
and the channel, $G_\ga(z)$, resolvents are
\begin{eqnarray}
G(z)&=&(z-H_0- V)^{-1}{N}
\equiv R(z){N},\\
G_\ga(z)&=&(z-H_0- { N_2^{\ga}}V_2^{\ga})^{-1}{ N}
\equiv R_\ga(z){ N}.
\end{eqnarray}
Note that $V^\dagger\neq V$ and $R(z)N\neq NR(z)$.
For the scattering problem it is convenient to define the in-medium
AGS operator $U_{\gb\ga}(z)$~\cite{bey96}
\begin{equation}
R(z)=\gd_{\gb\ga}R_\gb(z) + R_\gb(z) { U_{\gb\ga}(z)} R_\ga(z)
\end{equation}
that after some algebra leads to the in-medium AGS equation 
\begin{equation}
 U_{\gb\ga}(z)=\bar\gd_{\gb\ga}R_0(z)^{-1}+\sum_\gc
\bar\gd_{\gb\gc} 
N_2^\gc T_2^\gc(z) R_0(z) U_{\gc\ga}(z),
\end{equation}
where $\bar\gd_{\gb\ga}\equiv1-\gd_{\gb\ga}$.  The square of this AGS-operator
is directly linked to the differential cross section for the scattering
process $\ga\rightarrow\gb$, for all Fermi functions $f(i)\rightarrow 0$.
Hence the isolated three-body system is recovered. The driving kernel consists
of the two-body $t$-matrix derived in the same formalism, however given
earlier and known as Feynman-Galitskii (finite $T$) or Bethe-Goldstone ($T=0$)
equations~\cite{fet71,bet58}
\begin{eqnarray}\nonumber
T_2^\gamma(z) &=&   V_2^\gamma + 
 V_2^\gamma { N^\gc_2}R_0(z)  T_2^\gamma(z)\nonumber.
\end{eqnarray}
A numerical solution of the three-body break-up reaction relevant for the
chemical distribution in a heavy ion collision using a coupled Yamaguchi
potential has been given in Ref.~\cite{bey96}.

For the bound state problem it is convenient to introduce form factors
\begin{equation}
|F_\gb\rangle=\sum_\gc\bar\gd_{\gb\gc} { N_2^\gc}  V_2^\gc 
|\psi_{B_3}\rangle.
\end{equation}
Since the potential is non symmetric, the right and left eigenvectors are
different, although the bound state energies are the same. The eigenvectors
are explicitly needed in our solution of the four-body system. The respective
homogeneous AGS equations are given by
\begin{eqnarray}
|F_\ga\rangle
&=&\sum_\gb \bar\gd_{\ga\gb}  
{ N_2^\gb} T_2^\gb(B_3) R_{0}(B_3)|F_\gb\rangle,\nonumber\\
|\tilde F_\ga\rangle
&=&\sum_\gb \bar\gd_{\ga\gb} T_2^\gb(B_3) { N_2^\gb} R_{0}(B_3)
|\tilde F_\gb\rangle.\label{eqn:RL}
\end{eqnarray}

We now turn to the four-body problem in matter. In addition to having
different channels as for the three body system now the channels appear in
different partitions that makes the four-body problem even more involved. The
partitions of the four-body clusters are denoted by $\rho,\tau,\sigma,\dots$,
e.g., $\rho=(123)(4),(234)(1), \dots$ for $3+1$-type partitions, or
$\rho=(12)(43), (23)(41), \dots$ for $2+2$-type partitions.  The two-body
sub-channels are denoted by pair indices $\ga,\gb,\gc,\dots$, e.g. pairs
$(12)$, $(24)$,\dots The two- and three-body $t$-matrices have to be defined
with respect to the partitions that leads to additional indices.  A
convenient way to solve the four-body in-medium homogeneous AGS equation is
by introducing form factors
\begin{equation}
        |{\cal F}_\gb^\gs\rangle = \sum_\tau \bar\gd_{\gs\tau}
        \sum_\ga \bar\gd^\tau_{\gb\ga} R_0^{-1}(B_4) |\psi_{B_4}\rangle,
\end{equation}
where $\bar\gd^\tau_{\gb\ga}=\bar\gd_{\gb\ga}$, if $\gb,\ga\subset\tau$ and
$\bar\gd^\rho_{\gb\ga}=0$ otherwise and $|\psi_{B_4}\rangle$ is the
$\alpha$-particle in-medium wave function.
The homogeneous equations then read~\cite{Beyer:2000ds}
\begin{equation}
|{\cal F}^\gs_\gb\rangle=\sum_{\tau\gc} \bar\gd_{\gs\tau}
U^\tau_{\gb\gc}(B_4)  R_0(B_4) { N_2^\gc} 
T_2^\gc(B_4) 
R_0(B_4) |{\cal F}^\tau_\gc\rangle,
\end{equation}
where $\ga\subset\gs,\gc\subset\tau$.  A numerical solution of this equation
is rather complex. In order to reduce computational time, needed in particular
to handle the dependence on the medium, we introduce an energy dependent pole
expansion (EDPE) that has been proven useful in many applications involving
the $\alpha$-particle and is accurate enough for the present
purpose~\cite{EDPE}.  However, we have to generalize the original version of
the EDPE because of different right and left eigenvectors appearing for the
three-body subsystem and given in Eq.~(\ref{eqn:RL}) (for details
see~\cite{Beyer:2000ds}).

In the {\em two-body} sub-system the EDPE reads
\begin{eqnarray}
 T_\gc(z) &\simeq &\sum_n
 |\tilde\Gamma_{\gc n}(z)\rangle t_{\gc n}(z)\langle \Gamma_{\gc n}(z)|
 \nonumber\\
 &\simeq&\sum_n |\tilde g_{\gc n}\rangle t_{\gc n}(z)\langle g_{\gc n}|
 \nonumber\\
 &=&\sum_n
 {  N_2^\gc} |g_{\gc n}\rangle t_{\gc n}(z)\langle g_{\gc n}|.
 \label{eqn:Tsep2}
\end{eqnarray}
where we have chosen a Yamaguchi ansatz for the form factors for simplicity.
The last line shows the explicit dependence of the Pauli blocking factors.
Inserting this ansatz into the Feynman-Galitskii equation determines the
propagator $t_{\gc n}(z)$. In the {\em three-body} sub-system the EDPE
expansion reads
 \begin{eqnarray}
 \lefteqn{\langle g_{\gb m}(z)| R_0(z) U^\tau_{\gb\gc}(z)R_0(z)| 
 \tilde g_{\gc n}(z)\rangle
 \simeq}\nonumber\\
 && \sum_{t,\mu\nu} |\tilde\Gamma^{\tau t, \mu}_{\gb m}(z)\rangle
 t^{\tau t}_{\mu\nu}(z)\langle \Gamma^{\tau t, \nu}_{\gc n}(z)|.
 \label{eqn:pole3}
 \end{eqnarray}
 with the three-body EDPE functions 
 \begin{equation}
 |\tilde\Gamma^{\tau t, \mu}_{\gb m}(z)\rangle
 = \langle g_{\ga n}|R_0(z)| \tilde g_{\gb m}\rangle
 t_{\gb m}(B_3) |\tilde\Gamma^{\tau t, \mu}_{\gb m}\rangle,
 \end{equation}
 that we get from solving the following Sturmian equations
 \begin{eqnarray}
 \eta_{t,\mu}|\tilde\Gamma^{\tau t, \mu}_{\ga n}\rangle=
 \sum_{\gb m}
  \langle g_{\ga n}|R_0(B_3)| \tilde g_{\gb m}\rangle
 t_{\gb m}(B_3)|\tilde\Gamma^{\tau t, \mu}_{\gb m}\rangle
 \label{eqn:sturm}\\
 \eta_{t,\mu}|\Gamma^{\tau t, \mu}_{\ga n}\rangle=
 \sum_{\gb m}
  \langle \tilde g_{\ga n}|R_0(B_3)|  g_{\gb m}\rangle
 t_{\gb m}(B_3)|\Gamma^{\tau t, \mu}_{\gb m}\rangle
 \end{eqnarray}
 
 Inserting the EDPE into the homogeneous AGS equations allows us to redefine
 the form factors that are now operators depending on the coordinates of the
 $2+2$ or $3+1$ system, i.e.
\begin{equation} 
|\Gamma^{\gs s}_\nu\rangle 
= \sum_{\gb m}\langle\Gamma^{\gs s}_{\gb m,\nu}(B_4)|t_{\gb m}(B_4)
\langle g_{\gb m}(B_4)| R_0(B_4)|{\cal F}^\gs_\gb\rangle
\end{equation} 
and therefore the final homogeneous equation
\begin{eqnarray}
|\Gamma^{\gs s}_\mu\rangle& =
&\sum_{\tau t}\sum_{\nu\gk} \sum_{\gc n} \bar\gd_{\gs\tau}
\langle\Gamma^{\gs s, \nu}_{\gc n}(B_4)|t_{\gc n}(B_4)
|\tilde\Gamma^{\gs s, \mu}_{\gc n}(B_4)\rangle\;
\nonumber\\
&&\qquad\qquad\qquad\times 
t^{\tau t}_{\mu\gk}(B_4)\;|\Gamma^{\tau t}_\gk\rangle,
\label{eqn:coup4}
\end{eqnarray}
is an effective one-body equation with an effective potential ${\cal V}$ and
an effective resolvent ${\cal G}_0$ defined as
\begin{eqnarray}
{\cal V}^{\gs s,\tau t}_{\mu\nu}(z)
&= &\sum_{\gc n} \bar\gd_{\gs\tau}
\langle\Gamma^{\gs s, \mu}_{\gc n}(z)|t_{\gc n}(z)
|\tilde\Gamma^{\gs s, \nu}_{\gc n}(z)\rangle,
\label{eqn:pot4}\\
{\cal G}^{\gs s,\tau t}_{\mu\nu,0}(z)&=& t^{\tau t}_{\mu\nu}(z).
\end{eqnarray}

\section{Results}\label{sec:results} 

\subsection{Cluster properties}
\begin{figure}[t]
\begin{center}
\epsfig{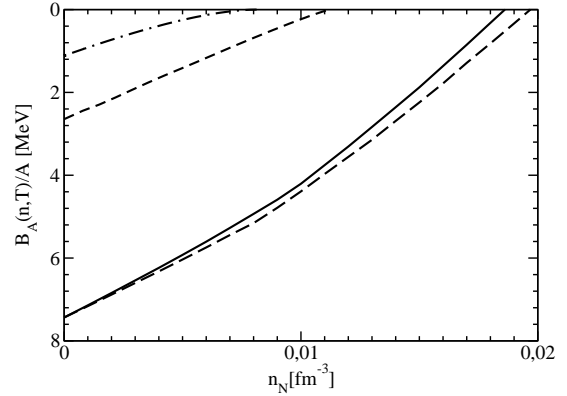}
\caption{\label{fig:Mott} Density dependence of the 
  binding energy per nucleon of deuterons (dash-dot), Triton (dashed), and
  $\alpha$-particle with Malfliet Tjon potential (solid), Yamaguchi potential
  (long-dashed) at $P_{\mathrm {c.m.}}=0$, as given in
  Ref.~\cite{Beyer:2000ds}.}
\end{center}
\end{figure}
\begin{figure}[t]
\begin{center}
\epsfig{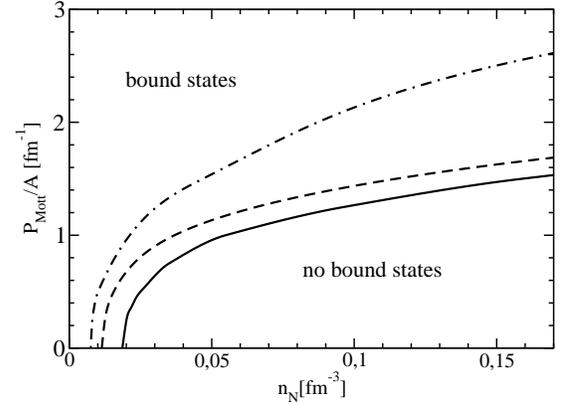}
\caption{\label{fig:MottV} Momenta per nucleon as a 
  function of the dissociation density for deuteron (dash-dot), triton (dash),
  $\alpha$-particle (solid). Bound states exist only above the respective
  lines.}
\end{center}
\end{figure}
\begin{figure}[t]
\begin{center}
\epsfig{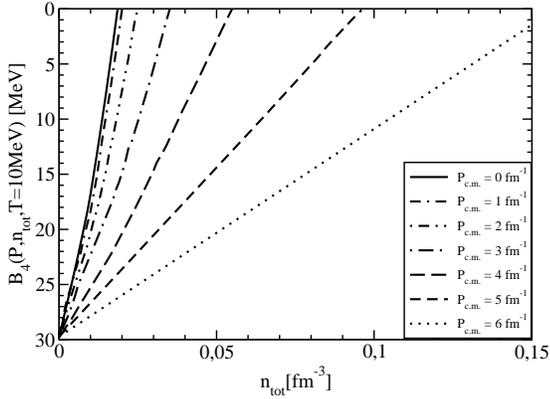}
\caption{\label{fig:AmottPT} Density dependence of the 
binding energy of the $\alpha$-particle for finite $P_{\mathrm {c.m.}}$
as indicated in the legend.}
\end{center}
\end{figure}

The binding energies of the few-nucleon systems depend on the chemical
potential $\mu$ or equivalently the density $n(\mu,T)$, the temperature $T$,
and c.m. momentum $P_{\mathrm {c.m.}}$. For the two-, three-, and four-nucleon
systems the binding energies are shown in Fig.~\ref{fig:Mott} for $T=10$ MeV
and $P_{\mathrm {c.m.}}=0$~\cite{Beyer:2000ds}. The line $B_A=0$ reflects the
respective continuum threshold. We mention here that the medium dependence of
the binding energies is rather similar for different two-body potentials,
although their results for the isolated system may be very different for the
few-nucleon systems considered. This has been mentioned earlier, but it can
also be seen from the two lines representing different potentials for the
$\alpha$-particle, i.e. Yamaguchi~\cite{yama} (long dashed) and Malfliet-Tjon
potential~\cite{MT} (solid) after renormalizing the binding energies to the
same value of the MTI-III potential. For $^3$He that is not shown in
Fig.~\ref{fig:Mott} the dissociation density is slightly smaller due
to the Coulomb force that has been evaluated perturbatively. However, for
asymmetric nuclear matter, e.g.  $N_p/N_n\simeq 0.72$ (for the
$^{129}$Xe$+~^{119}$Sn reaction) this effect is compensated~\cite{mat00}.

Because of the medium the pole of the bound state moves to the continuum
threshold as seen in Fig.~\ref{fig:Mott}. The bound state vanishes because of
the quasi-particle nature of the cluster. However, investigating the zeros of
the two-body Jost function we found earlier that the quasi deuteron
``survives'' as a virtual bound state with different energies, depending on
the densities above the dissociation line~\cite{Beyer:2001nc}. This is a
similar state as the virtual $^1S_0$ nucleon nucleon state at 70 keV. This
means that the quasi deuteron retains its quasi-particle nature of being an
infinitely long living state and does not become a resonance. Only going
beyond the quasi-particle picture the deuteron spectral function will aquire
an imaginary part and hence the deuteron becomes a state of finite life time
in the medium. This is due to break-up processes in nuclear matter, the
simplest one being the three-nucleon reaction~\cite{bey96}. Such an
investigation for three- and four-body system is also technically involved and
still needs to be done.

Also a possible appearance of Efimov states related to $B\rightarrow 0$ of the
sub-system needs further investigation~\cite{efimov}.  Since the Efimov states
are 'excited' states, e.g.  for the three-body system they are close to the
$2+1$ threshold, their blocking may be smaller since the wave functions
contain higher momentum components, hence the slope of their dependence on the
density is flatter.  Also, as seen from Fig.~\ref{fig:Mott}, the slope of the
binding energies as a function of densities for the larger clusters is
steeper.  On the other hand the sub-system is not at rest in the larger
cluster, hence the binding energy changes, as we will show in the next
paragraph, and therefore a careful analysis is needed that would go beyond the
present scope of the paper. Hence so far no conclusion can be drawn for the
appearance of Efimov states, but it is an important issue since Efimov states
might effect the equation of state for clustering Fermi systems.

For a finite c.m. momentum relative to the medium (at rest) the influence of
the medium is weaker, as less components of the wave functions are blocked by
the Fermi sea. For deuteron~\cite{schnell} and triton~\cite{Beyer:1999zx} this
has been given in earlier references, see also references therein. In
Fig.~\ref{fig:AmottPT} we give the results for the $\alpha$-particle. Note
again that the medium effects do not change the elementary property of an
$\alpha$-particle, however, after introducing effective degrees of freedom,
the $\alpha$-particle and any other cluster considered here consists of
quasi-nucleons and not elementary nucleons.  Besides the change of nucleon self
energy also the binding energy of the cluster is changed and hence the
clusters can be viewed as quasi-deuterons, quasi-tritons, quasi-$\alpha$'s,
etc., i.e. clusters with the respective self-energy corrections.

For a given temperature, here we chose $T=10$ MeV, 
the momentum of dissociation $P_\mathrm{dis}$ is defined by the condition
\begin{equation}
B(n_{\mathrm {dis}},T,P_\mathrm{dis})=0,
\end{equation}
i.e., the density $n_{\mathrm {dis}}$ and the momentum $P_\mathrm{dis}$ where
binding of the nucleons is lost. For a system of atoms and ions this scenario
can be related to a transition of an isolator to a conductor, since electrons
can move away from the ion because of the dissociation property. The
respective dissociation lines for deuterons, tritons/$^3$He, and
$\alpha$-particles are shown in Fig.~\ref{fig:MottV}. The momentum is
normalized to the number of nucleons in the cluster that is identical to the
velocity of the cluster on the dissociation line.

\subsection{Cluster distribution}
\begin{figure}[t]
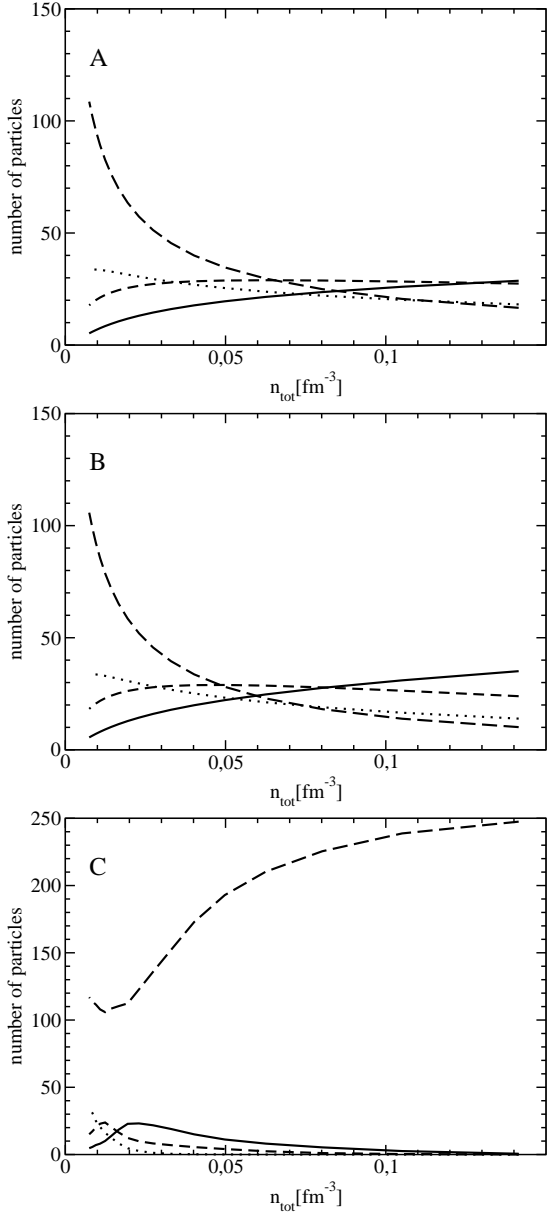

\begin{center}
\epsfig{figure=zahl_ideal.eps,width=0.4\textwidth}
\epsfig{figure=zahl_quasi.eps,width=0.4\textwidth}
\epsfig{figure=zahl_xb.eps,width=0.4\textwidth}
\caption{\label{fig:dichte} Numbers of  nucleon (long dashed), deuteron
  (dotted), triton ($^3$He) (dashed), $^4$He (solid) as a function of the
  total density $n_{\rm tot}$ at $T=10$ MeV. Total number of nucleons 250. A)
  ideal system B) quasi particle system, B) full calculation with dissociation. }
\end{center}
\end{figure}

We now consider the composition of nuclear matter in equilibrium at a
temperature of $T=10$ MeV. To this end we assume that nuclear matter is
composed of nucleons, deuterons, tritons, $^3$He, and $\alpha$-particles.
Larger clusters are presently not considered.  We investigate three scenarios:
\begin{enumerate}
\item[A.] gas of nucleons and nuclei with all properties of the isolated
  systems retained (no change due to medium),
\item[B.] a gas  of quasi-nucleons that contain self energy corrections, the
  clusters consist of quasi-nucleons but the interaction is without Pauli
  blocking, hence the binding energy will not change, and
\item[C.] a gas of quasi-nucleons with Pauli blocking in the interaction and
  therefore the clusters are treated as quasi-nuclei that include the self
  energy corrections on the cluster level and dissociation.
\end{enumerate}
In the ideal situation (case A), the components retain their properties, i.e. all
particle masses stay the same. The composition of the system is driven by the
law of mass action, i.e. the equilibrium distribution functions of nuclei
(consisting of $A$ nucleons) with mass $m_A=Am_N-B_A$ are given by
\begin{equation}
f_A(p)=\left\{\exp[\gb(p^2/2m_A-B_A-\mu_A)]+\epsilon\right\}^{-1}.
\end{equation}

The composition of the system, i.e., the number of particles as a function of
the total density, for $T=10$ MeV is given in Fig.~\ref{fig:dichte}A. The
density is accumulated by the more massive clusters the larger the total
densities gets. The freeze-out distribution could be read of at
$n_\mathrm{tot}=0.085\dots0.034$ fm$^{-3}$~\cite{Borderie:1995sk}.  
Fig.~\ref{fig:dichte}B refers to the result of the quasi-particle
approximation (case B) for nucleons, instead of using ideal nucleons. In this
case the medium effects are taken into account in Hartree-Fock approximation
on the single particle level. This results in different self energies for the
nucleon and in turn the mass of nuclei changes accordingly, see
Eq.~(\ref{eqn:quasi})
\begin{equation}
\varepsilon(k) \simeq  k^2/2m^{\mathrm {eff}} + \Sigma(0)
\label{eqn:self}
\end{equation}
The right hand side of Eq.~(\ref{eqn:self}), known as effective mass
approximation, is valid for the rather low momenta and densities considered,
hence $m^{\mathrm {eff}}=m_N(\mu,T)$ approximately independent of the momentum
$k$.  The distribution functions change to
\begin{equation}
f_A(p)=\{\exp[\gb(p^2/2m_{A}^{\mathrm{eff}}-B_A
-\mu_A^{\mathrm{eff}})]+\epsilon\}^{-1}
\end{equation}
and now $m_{A}^{\mathrm{eff}}=Am^{\mathrm{eff}}-B_A$. In chemical equilibrium
$\mu_A^{\mathrm{eff}}=A\mu^{\mathrm{eff}}$, where
$\mu^{\mathrm{eff}}=\mu-\Sigma(0)$.  The results are rather close to the ideal
gas case, because the change of the self energy of the cluster due to the
binding energy of the bound nuclei is not taken into account. Differences
appear at larger densities.

The situation changes drastically for case C, if the change of the binding
energy as discussed above is taken into account,
\begin{equation}
B_A\rightarrow B_A(p,T,\mu)\equiv B_A^{\mathrm{eff}}.
\end{equation}
This, however, needs a solution of few-body in-medium equations as given in
the previous section. The effects induced by this change in binding energy is
shown in Fig.~\ref{fig:dichte}C.  The equation for the
density of the cluster changes, because the bound state exists only above the
momentum of dissociation as shown in Fig.~\ref{fig:MottV}. The definition of
the density changes accordingly, see Eq.~(\ref{eqn:dens})
\begin{equation}
n^\mathrm{b}_A(\mu,T) = (2S+1)(2I+1)\sum\limits_{p>p_{\mathrm{dis}}} f_A(p).
\end{equation}
Also the distribution function is different from the previous definition,
since now the change of the binding energy has to be taken into account.
\begin{equation}
f_A(p)=\{\exp[\gb(p^2/2m_A^{\mathrm{eff}}-B_A^{\mathrm{eff}}
-\mu_A^{\mathrm{eff}})]+\epsilon\}^{-1}
\end{equation}
where now $m_A^{\mathrm{eff}}=Am_N(\mu,T)-B_A^{\mathrm{eff}}$ and 
$B_A^{\mathrm{eff}}$ is given in Fig.~\ref{fig:AmottPT} for the
$\alpha$-particle, for the three-body case in~\cite{Beyer:1999zx} and for the 
deuteron in~\cite{schnell}, and Refs. therein.

The change between the ideal (or quasi-particle) picture and the full
calculation that includes self energy corrections and dissociation of the
clusters appears quite decisive. Whereas for rigid nuclei the number of
heavy particles is much higher than the number of light particles, this is
different, if the dissociation is taken into account. The geometrical
interpretation of dissociation is because less low momentum components (large
distance components) are available for the formation of a bound state. If the
momentum is higher, the nucleus moves out of the Fermi sphere of the
surrounding matter and the particle becomes more stable.  The fact that no
bound states are possible doesn't mean that there are no correlations.
Previously we found by analyzing the Jost function of the deuteron that as the
deuteron moves towards lower binding energies and eventually crosses the
continuum line ($B_2=0$) it exists as a virtual bound state on the unphysical
energy sheet~\cite{Beyer:2001nc}. In its turn this means that particular
correlations in the continuum can form deuterons, if the density becomes low
enough (for a given temperature).  A similar study for three- and four-body
states still needs to be done.  On the other hand scattering states are
infinitely extended, in contrast to bound states.  So neglecting those
few-body correlations related to scattering states may not be the worst
approximation to start with.

\begin{center}
\begin{figure*}[t]
\begin{center}
\epsfig{figure=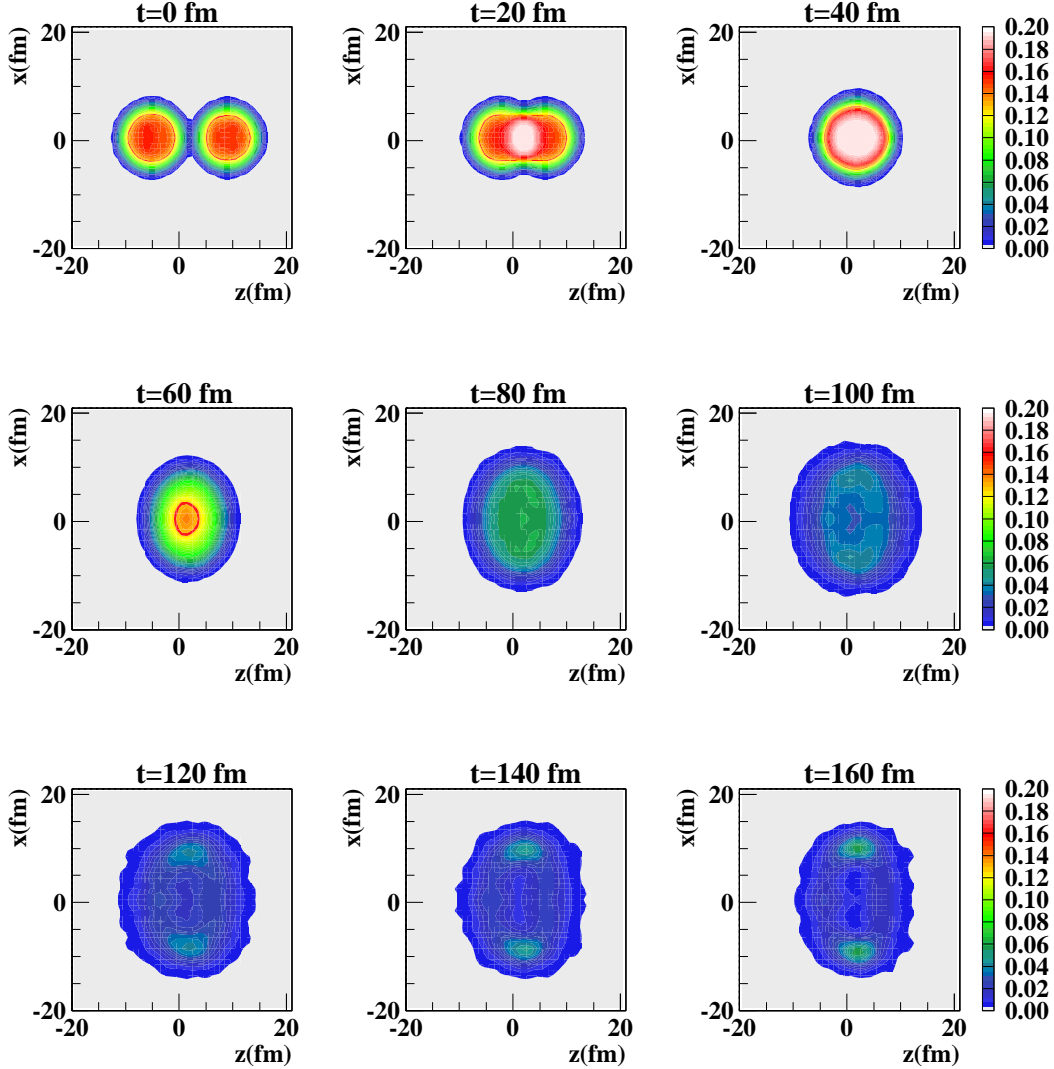,width=0.8\textwidth}
\caption{\label{fig:dens} Evolution of the density in the $xz$-plane as
provided by a BUU simulation~\cite{ckuhrts} of a central collision
 of Xe on Sn. }
\end{center}
\end{figure*}
\end{center}

To give an example for possible consequences of this finding, we investigate
the multiplicities of light fragments in a heavy ion collision. We investigate
conditions close to the INDRA experiment Xe on Sn at $50A$ MeV~\cite{INDRA00}
studied earlier within the context of a BUU simulation~\cite{Beyer:1999xv}.
We focus on the final stage of the collision and for simplicity assume a
homogeneous temperature of $T=10$ MeV that might slightly be too high for a
quantitative comparison, but is still reasonable.  Also, we use symmetric
nuclear matter that is not achieved in the experiment mentioned. However, the
basic effects that we focus on in this paper are not changed: We found
previously that asymmetric nuclear matter (using the asymmetry induced by the
experiment Xe on Sn) has very little effect on the dissociation of $^3$He and
triton.  Quantitatively, this effect is in the same order of magnitude as the
Coloumb correction~\cite{mat00}.

\begin{figure}[t]
\begin{center}
\epsfig{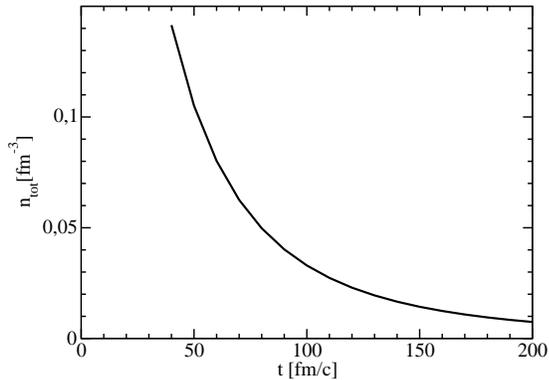}
\caption{\label{fig:evolu} Evolution of total density in a uniform model.
Parameters are chosen as explained in the text. }
\end{center}
\end{figure}

The BUU simulation of the central collision of Xe on Sn at $50A$ MeV provides
a realistic nuclear density distribution for the INDRA experiment.  A cut
through the $xz$-plane is shown in Fig.~\ref{fig:dens}~\cite{ckuhrts}.  To
simplify and model this density distribution we assume a homogeneous spherical
distribution of radius $R$ that approximately matches the size of the
simulation at about 40 fm/c ($R=7.5$ fm) and about 140 fm/c ($R=20$ fm) after
the collision. The radial change is assumed to be linear.  The resulting
change of the local density with respect to time for this simple expanding
fire ball is shown in Fig.~\ref{fig:evolu}.

\begin{figure}[t]
\begin{center}
\epsfig{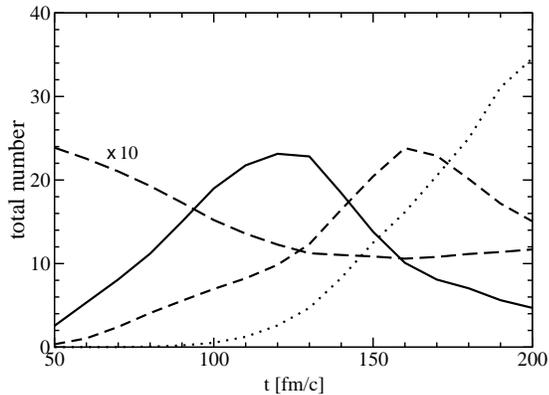}
\caption{\label{fig:zahl} Total numbers of nucleons (long-dashed$\times 10$), 
deuteron (dots), triton/$^3$He (dashed), 
and $\alpha$ particles (solid) as a function of collision time at $T=10$ MeV.}
\end{center}
\end{figure}

The multiplicities of nucleons, deuterons, $^3$He/$^3$H, and
$\alpha$-particles for the densities evolution of Fig.~\ref{fig:evolu} and a
temperature of $T=10$ MeV for a total number of nucleons of 250 is given in
Fig.~\ref{fig:zahl}. This is the main result of the present calculation.

\subsection{Discussion}
At around $t=100$ fm/c the number of $\alpha$-particles is much larger than
the number of other clusters. This time, which (approximately) corresponds to
about 3-4 times the initial volume, can be considered as freeze-out time
(related to the freeze-out volume~\cite{INDRA02}). Hence the multiplicities at
this time should be significantly correlated with the experimentally observed
ones. Indeed, the total multiplicities of $\alpha$-particles in the above
mentioned INDRA experiments Xe+Sn is larger than those of the lighter
clusters~\cite{INDRA02}.

In contrast to the full calculation, the other equilibrium scenarios discussed
above have the opposite ordering of multiplicities which can be concluded from
Fig.~\ref{fig:dichte} for $n_{\mathrm{tot}}\simeq 0.04 \mathrm{fm}^{-3}$ and
does not reflect the experimental finding for the large excess of
$\alpha$-particles.  We argue that the enhancement of $\alpha$-particles is
also related to the fact that the $\alpha$-particle is more stable in low
density nuclear matter than the other light clusters.  Our findings of the
ordering of multiplicities for the light fragments obtained via a microscopic
approach and including dissociation might provide a natural explanation to the
excess of $\alpha$-particles.  However, for a throughout comparison with
experimental data several other aspects have to be taken into account as
mentioned in~\cite{INDRA02,mek78}. However, some of them need a major effort
while going beyond the model of an ideal gas of components.

Concerning larger (light) clusters than the ones considered so far, note that
they are weaker bound than the $\alpha$-particle. Therefore they should be
less stable in medium.  Hence, at freeze-out their multiplicity should be
smaller than that of the $\alpha$-particle.  However, a more quantitative
analysis is certainly needed. Within the equilibrium scenario the most stable
nucleus might be Fe. However, little is known about the properties of Fe at
finite temperature. At low densities one might expect a linear dependence of
the binding energy (perturbative theory), however the calculation for the
$\alpha$-particle clearly shows that this might not be valid for a stronger
bound system. On the other hand during (central) heavy ion collision the
dynamical generation of heavy nuclei needs time due to many (binary)
collisions in the system and such heavy elements might not recombine at all in
such a evaporation scenario~\cite{Borderie:1995xc}.

A microscopic treatment of more complex (light) nuclei could be achieved,
e.g., along the lines of~\cite{sofFB}. So far, larger and heavier clusters are
''hidden'' in the large number of nucleons present. While those are rather
easy to incorporate in the ideal gas picture, a calculation including medium
dependence such as self energy corrections and Pauli blocking is more
elaborate.

\section{Conclusion}
\label{sec:con}
We have shown that a systematic microscopic calculation provides strong
changes in the equilibrium composition of clusters in nuclear matter. The
changes are strong enough to invert the ordering of multiplicities at
freeze-out compared to the ideal case. Therefore an explanation of
experimental results in terms of a microscopic picture with realistic nucleon
nucleon forces evaluated in an equilibrium scenario might be possible.  Note
that a detailed comparison of this approach to the experimental data as, e.g.,
given by the INDRA collaboration for the SMM~\cite{INDRA00,INDRA02}, needs
much further investigation and has to be postponed to a future communication.


{\em Acknowledgment:} We thank Christiane Kuhrts for providing us with
Fig.~\ref{fig:dens}~\cite{ckuhrts} and Gerd R\"opke for discussions. MB
acknowledges the warm hospitality of the IPN Groupe Th\'eorie and the Physics
Department of UNISA during longer research stays. Work supported by Deutsche
For\-schungs\-ge\-mein\-schaft BE 1092/7.


\begin{thebibliography}{99}
\bibitem{dan91} P. Danielewicz and G.F. Bertsch, Nucl. Phys. {\bf A533}, 712
  (1991).
\bibitem{dan92} P. Danielewicz and Q. Pan, Phys. Rev. C {\bf 46}, 2002
  (1992). 
\bibitem{sto86} H. St\"ocker and W. Greiner, Phys. Rep. {\bf 137}, 277 (1986).
\bibitem{fuc95} C. Fuchs and H.H. Wolter, Nucl. Phys. {\bf A589}, 732 (1995).
\bibitem{Beyer:1999xv}
M.~Beyer, C.~Kuhrts, G.~R\"opke and P.~D.~Danielewicz,
Phys. Rev. C {\bf 63}, 034605 (2001).
\bibitem{aic91} J. Aichelin, Phys. Rep. {\bf 202}, 233 (1991).
\bibitem{pei92} G. Peilert {\em et al.}, Phys. Rev. C {\bf 46}, 1457 (1992).
 \bibitem{neb99} R. Nebauer and J. Aichelin, Nucl. Phys. {\bf A650}, 65 (1999);
INDRA Collaboration, R. Nebauer {\em et al.}, Nucl. Phys. {A658}, 67 (1999).
\bibitem{INDRA00} INDRA Collaboration, D. Gorio {\em et al.}, Eur. Phys. J. A
  {\bf 7}, 245 (2000), and references therein.
\bibitem{INDRA02} INDRA Collaboration, 
S. Hudan {\em et al.}, Phys. Rev. {\bf C 67}, 064613 (2003).
%
\bibitem{Borderie:1995sk}
INDRA Collaboration, B.~Borderie {\it et al.},
Phys.\ Lett.\ B {\bf 388}, 224 (1996).
%
\bibitem{Borderie:1995xc}
INDRA Collaboration, B.~Borderie {\it et al.},
Phys.\ Lett.\ B {\bf 353}, 27 (1995).
\bibitem{mek78} A.Z. Mekjan, Phys. Rev. {\bf C 17}, 1051 (1978).
\bibitem{tri94} R.K. Tripathi and L.W. Townsend, Phys. Rev. {\bf C50}, R7 (1994).
\bibitem{bon95} J.P. Bondorf, A.S. Botvina, A.S. Iljinov, I.N. Mishustin, and
  K. Sneppen, Phys. Rept. {\bf 257}, 133 (1995).
\bibitem{Beyer:1999zx}
M.~Beyer, W.~Schadow, C.~Kuhrts and G.~R\"opke,
Phys.\ Rev.\ C {\bf 60}, 034004 (1999).
\bibitem{Beyer:2000ds}
M.~Beyer, S.~A.~Sofianos, C.~Kuhrts, G.~R\"opke and P.~Schuck,
Phys.\ Lett.\ {\bf B488}, 247 (2000).
 \bibitem{ckuhrts} Chr. Kuhrts, {\em PhD thesis: ``Deuteron production in
 heavy ion reactions''} Rostock 2000.
\bibitem{fet71} For a textbook treatment see, e.g., 
L.P. Kadanoff, G. Baym, {\em Quantum Theory of
    Many-Particle Systems} (Mc Graw-Hill, New York, 1962);
 A.L. Fetter, J.D. Walecka, {\em Quantum Theory of
    Many-Particle Systems}, (Mc Graw-Hill, New York, 1971).
\bibitem{duk98} see e.g., J. Dukelsky,  G. R\"opke, and P. Schuck, Nucl. Phys. {\bf
    A 628}, 17 (1998).
\bibitem{bey96} M. Beyer, G. R\"opke, and A. Sedrakian, Phys. Let. B {\bf 376},
    7 (1996).
\bibitem{bet58} H.A. Bethe and J. Goldstone, Proc. R. Soc. {\bf A 238}, 551 (1957).
\bibitem{eic68} J. Eichler, T. Marumori, and K. Takada,
  Prog. Theor. Phys. {\bf 40}, 60 (1968).
\bibitem{sch73} P. Schuck, F. Villars, and P. Ring, Nucl. Phys. {\bf A 208},
  302 (1973).
\bibitem{schulz}
M. Schmidt, G. R\"opke, H. Schulz, Ann. Phys. (N.Y.) {bf 202}, 57 (1990).
\bibitem{alt67} E.O. Alt, P.
  Grassberger, W. Sandhas, Nucl. Phys. {\bf B 2} (1967) 167.
\bibitem{san74} W. Sandhas, Acta Physica Austriaca, Suppl. {\bf XIII}, 
  679 (1974).
\bibitem{alt72} E.O. Alt, P. Grassberger and W. Sandhas, Report
  E4-6688, JINR, Dunba 1972 and in {\em Few particle problems in the
    nuclear interaction} eds. I. Slaus et al. (North Holland,
  Amsterdam 1972) p. 299.
\bibitem{san75} W. Sandhas, Czech. J. Phys. {\bf B 25}, 251 (1975).
\bibitem{EDPE} S. Sofianos, N.J. McGurk, and H. Fiedeldey, Nucl. Phys.  {\bf A318}, 295
(1979); S.A. Sofianos, H.  Fiedeldey, H. Haberzettl, and W. Sandhas
Phys. Rev. {\bf C 26}, 228 (1982).
\bibitem{yama}
Y. Yamaguchi, Phys. Rev. {\bf 95}, 1628 (1954).
\bibitem{MT}
R.A. Malfliet and J. Tjon, Nucl. Phys. {bf A 127}, 161 (1969).
\bibitem{mat00} S. Mattiello, {\em diploma thesis: ``Production of three-body
    clusters in asymetric matter'' (in Italian)}, U. Trento 2000 (unpublished)''
\bibitem{Beyer:2001nc}
M.~Beyer and S.~A.~Sofianos,
J.\ Phys.\ G {\bf 27}, 2081 (2001).
\bibitem{efimov} Efimov, V. N., Yad. Fiz {\bf 12}, 1080 (1970)
[  Sov. J. Nuc. Phys.  {\bf 12}, 589 (1971)].
\bibitem{schnell} A. Schnell, {\em PhD thesis}, U. Rostock 1999.
\bibitem{sofFB} S.A. Sofianos, 17th International Few-Body Conference  (2003).
 
\end{thebibliography}
\end{document}